\documentclass[11pt,twoside]{article}
\usepackage{asp2010}

\resetcounters

\begin{document}

\title{HD\,76431 - An evolved hot subdwarf with variable magnetic field?}
\author{G. Chountonov$^1$ and S. Geier$^2$
\affil{$^1$Special Astrophysical Observatory, Nizhnij Arkhyz, Zelenchukskiy region, Karachai-Cherkessian Republic, Russia 369167}
\affil{$^2$Dr. Karl Remeis-Observatory \& ECAP, Astronomical Institute,
Friedrich-Alexander University Erlangen-Nuremberg, Sternwartstr. 7, D 96049 Bamberg, Germany}}

\begin{abstract}
We measured the magnetic field of the bright, evolved hot subdwarf HD\,76431 by means of high-resolution spectropolarimetry. In contrast to previous measurements we were not able to detect a significant magnetic field. We discuss the possibility that this field may be variable. Our search for a possible companion star to HD\,76431 led to inconclusive results.

\end{abstract}

\section{Introduction}

\begin{figure}
\begin{center}
\includegraphics[width=10cm]{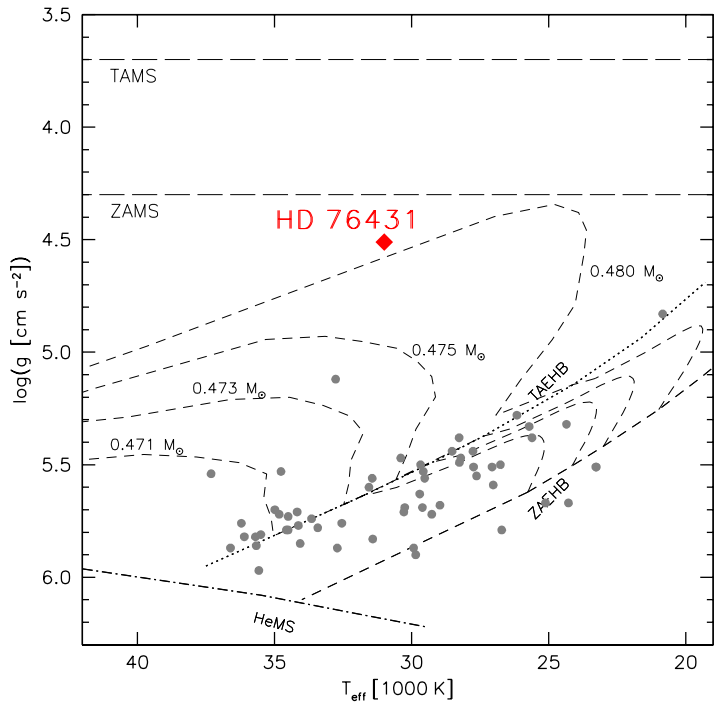}
\label{fig:tefflogg}
\caption{$T_{\rm eff}-\log{g}$ diagram. The grey circles mark sdBs from the SPY project \citep{lisker05}. EHB evolutionary tracks are taken from \citet{dorman93}. The location of the main sequence is indicated by the long-dashed horizontal lines. The atmospheric parameters of HD\,76431 are taken from Ramspeck et al. (2001).}
\end{center}
\end{figure}

HD\,76431 (PG\,0853+019, $m_{\rm B}=8.9\,{\rm mag}$) is a bright B-type star located at high Galactic latitude. According to its atmospheric parameters \citep[$T_{\rm eff}=31\,000\,{\rm K}$, $\log{g}=4.51; $][]{ramspeck01} it is located below the main sequence in the $T_{\rm eff}$-$\log{g}$-diagram (see Fig.~\ref{fig:tefflogg}). \citet{ramspeck01} measured a subsolar He abundance ($\log{y}=-1.51$) and concluded that HD\,76431 is most likely a post-EHB object.

The role of magnetic fields in stellar evolution remains unclear. One way to address this issue is the analysis of the magnetic properties of evolved stars. While the strong MG-fields of magnetic white dwarfs are easily detectable due to the prominent Zeeman splitting of their spectral lines, weaker fields can only be detected with spectropolarimetry. Such fields of the order of a few kG have been detected in white dwarfs and central stars of planetary nebulae.
 
A first attempt to determine the magnetic fields of bright hot subdwarf stars using this method was made by \citet{elkin98}, who also targeted HD\,76431, but did not detect a magnetic field ($B_{\rm e}=-50\pm130\,{\rm G}$). \citet{otoole05} on the other hand reported the detection of a magnetic field in HD\,76431 with a mean longitudinal field strength of $-1096\pm91\,{\rm G}$ using the medium resolution instrument FORS1 mounted at the ESO-VLT. Here we present the results of a spectropolarimetric monitoring campaign of this star using high-resolution data.

\section{Data Analysis}

\begin{figure}
\begin{center}
\includegraphics[angle=-90,width=10cm]{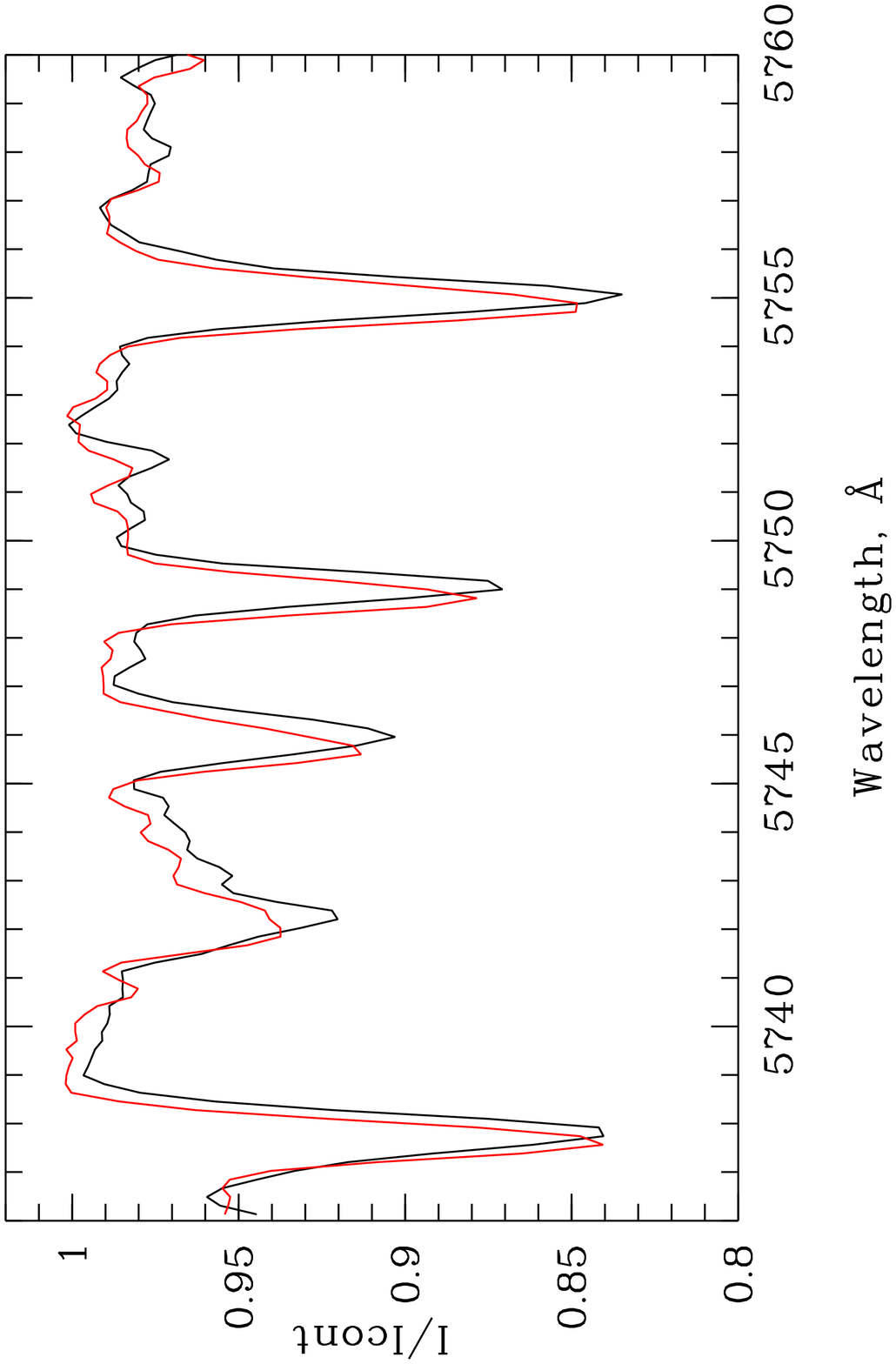}
\includegraphics[angle=-90,width=10cm]{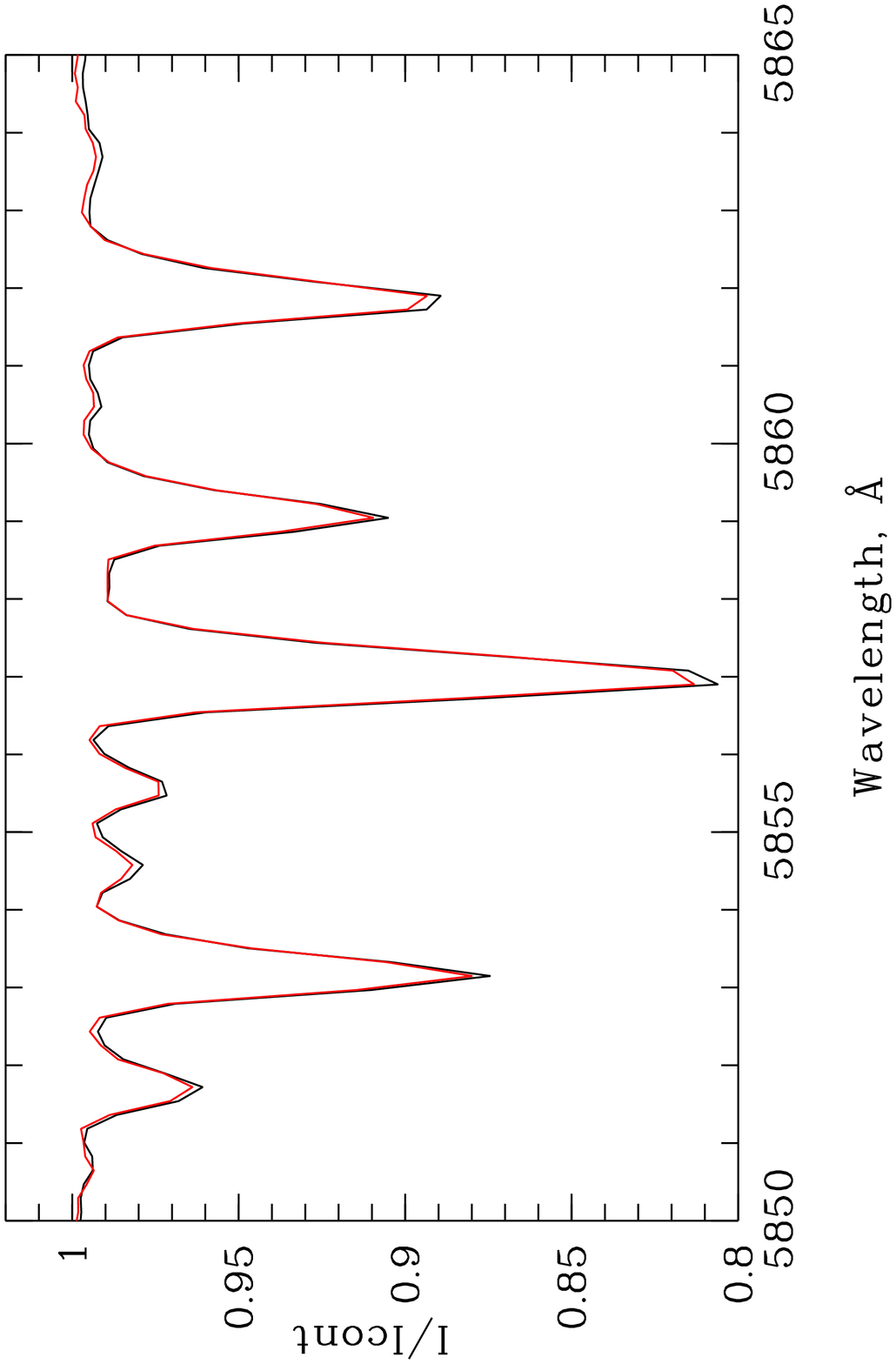}
\label{fig:53cam}
\caption{Upper panel: Spectrum fragment of the magnetic field standard 53\,Cam in the two polarizations (Phase=0.85, $B_{\rm e}=-4020\pm220\,{\rm G}$). Lower panel: Spectrum fragment of the non-magnetic star Procyon in the two polarizations ($B_{\rm e}=-150\pm80\,{\rm G}$).}
\end{center}
\end{figure}

Magnetic field measurements were performed with an analyser of circular polarization and dual-slicer (14 slices) at a spectral resolution of $15\,000$ on the Main stellar spectrograph mounted at the 6m telescope of the Special Astrophysical Observatory (SAO) in Northern Caucasus. Eight spectra of HD\,76431 have been obtained with the quarter-wave plate rotating at $90^{\rm \circ}$ and back with individual exposure times of $15\,{\rm min}$. Spectra of standard stars have been observed with the same setup. Helium as well as metal lines have been used to determine the longitudinal magnetic field strengths. Fig.~\ref{fig:53cam} (upper panel) shows a spectrum of the magnetic standard star 53\,Cam in the two polarizations. The shift due to the measured field is clearly visible. Fig.~\ref{fig:53cam} (lower panel) shows a similar plot of the non-magnetic star Procyon. Both plots demonstrate the accuracy of our measurements.

\section{Results}

\begin{figure}
\begin{center}
\includegraphics[angle=-90,width=10cm]{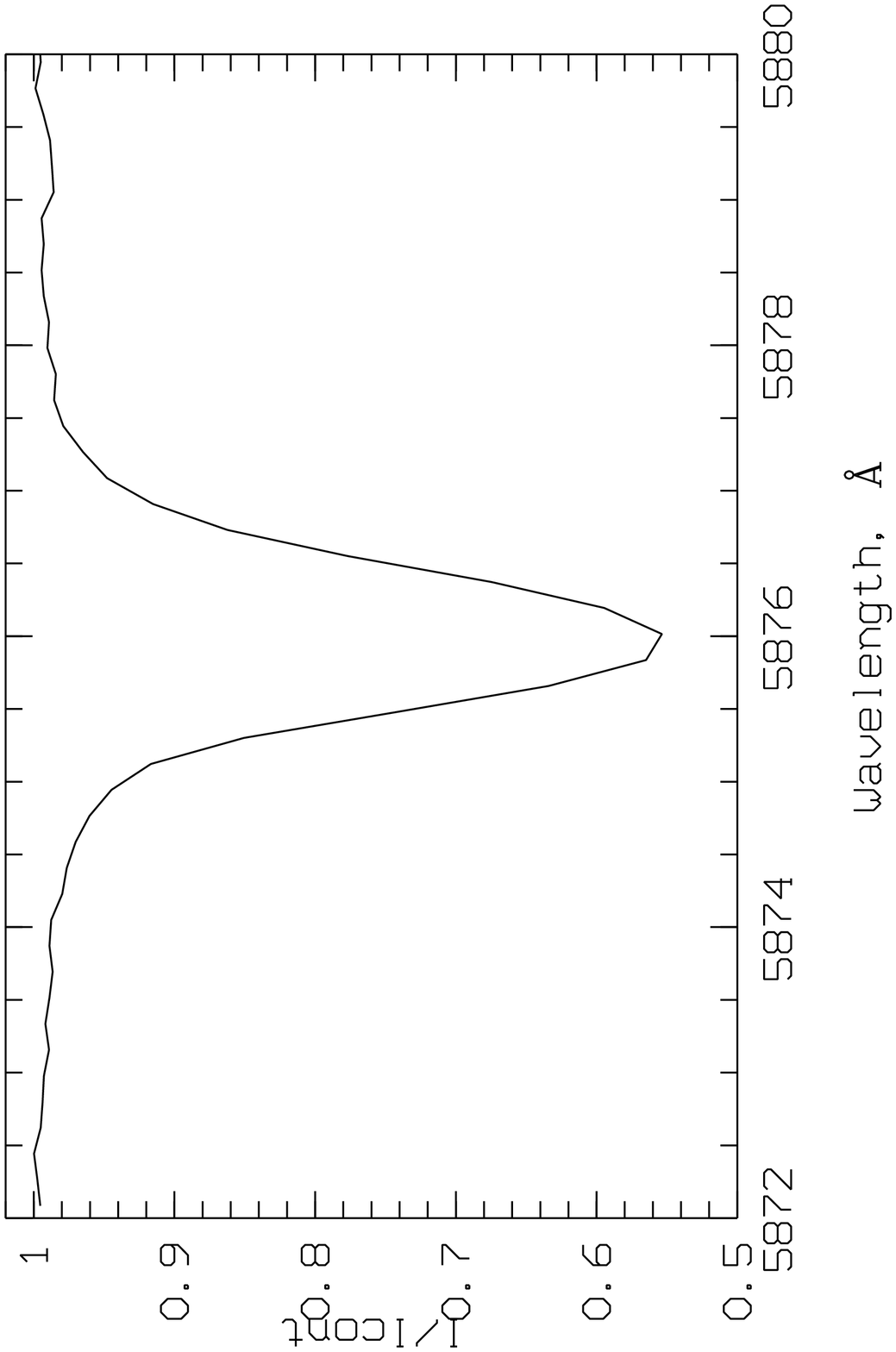}
\includegraphics[angle=-90,width=10cm]{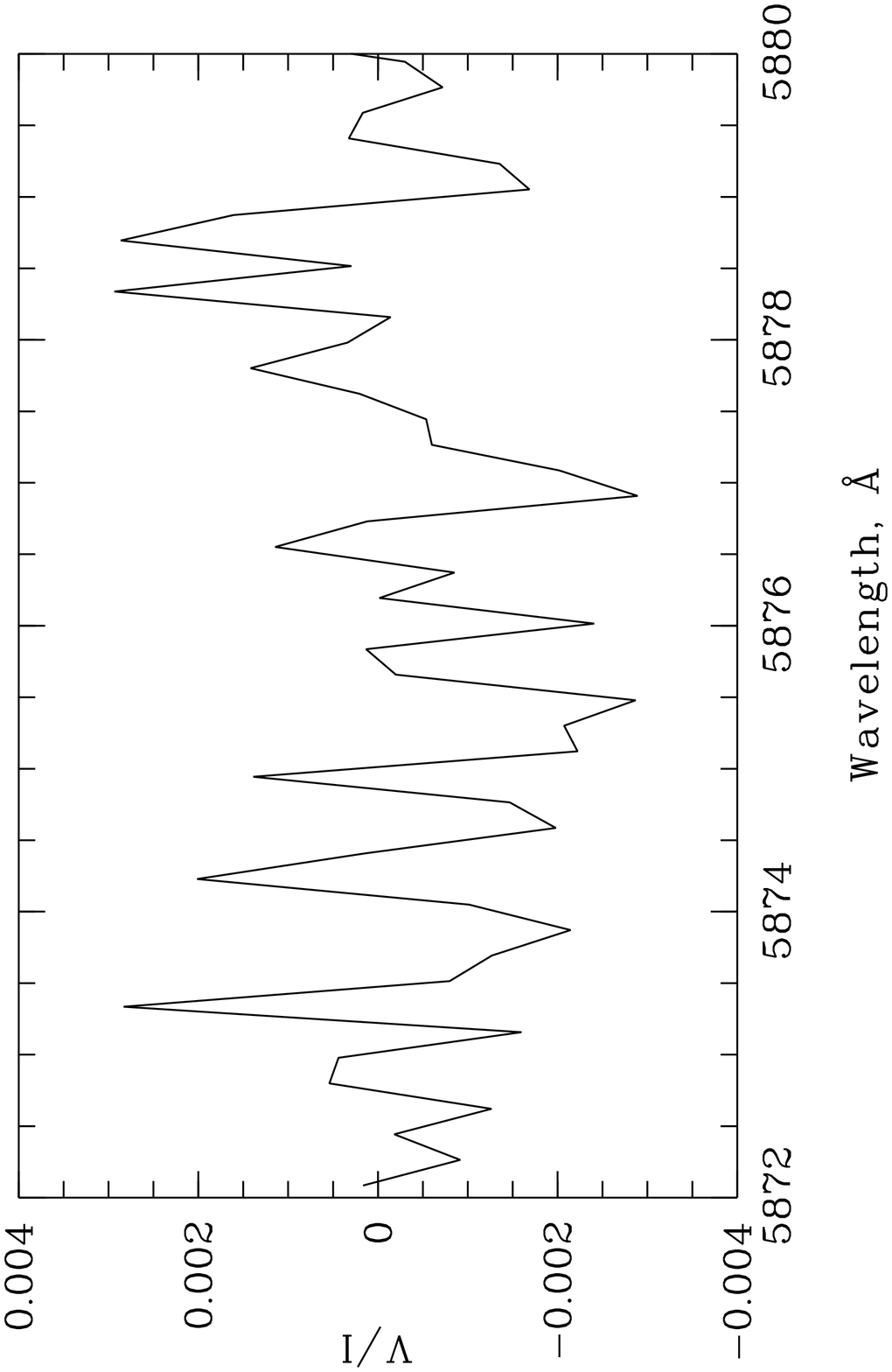}
\caption{Upper panel: SAO spectrum of HD\,76431 around the He {\sc i} $5876\,{\rm \AA}$ line. Lower panel: Circular polarization plotted against wavelength in the same spectral region. The field strength limit derived from these data is $-61\pm120\,{\rm G}$.}
\label{fig:hd}
\end{center}
\end{figure}

The results of our monitoring campaign are given in the table below. In order to determine the longitudinal magnetic fields we used helium lines as well as sharp metal lines in two different wavelength ranges. We were not able to detect a significant magnetic field. Our upper limits are of the order of $\simeq100-200\,{\rm G}$. An example is shown in Fig.~\ref{fig:hd}. 

\begin{table}
\begin{center}
\caption{Magnetic field measurements of HD\,76431}
\begin{tabular}{llrrr}
\hline
\noalign{\smallskip}
Date [JD] & Range [${\rm \AA}$] & & $n_{\rm lines}$ & $B_{\rm e}\,{[G]}$ \\
\noalign{\smallskip}
\hline
\noalign{\smallskip}
$2453746.49$ & $5600-5900$ & He\,I     & $1$  & $-61\pm120$ \\
             &             & Metal  & $9$  & $+11\pm65$ \\
$2453747.48$ & $5600-5900$ & He\,I     & $1$  & $-52\pm110$ \\
             &             & Metal  & $10$ & $+47\pm70$ \\
$2453748.55$ & $5600-5900$ & He\,I     & $1$  & $+41\pm110$ \\
             &             & Metal  & $11$ & $+132\pm70$ \\
$2453749.47$ & $5600-5900$ & He\,I     & $1$  & $-128\pm120$ \\
             &             & Metal  & $10$ & $-42\pm60$ \\
$2455196.52$ & $4360-4610$ & Metal  & $5$  & $+236\pm230$ \\
$2455252.41$ & $4360-4610$ & He\,I   & $2$  & $-145\pm180$ \\
             &             & Si\,III & $3$  & $-4\pm240$ \\
             &             & Metal  & $19$ & $+117\pm70$ \\
$2455253.41$ & $4360-4610$ & He\,I   & $2$  & $-123\pm170$ \\
             &             & Si\,III & $3$  & $-93\pm140$ \\
             &             & Metal  & $19$ & $-12\pm60$ \\
$2455254.37$ & $4360-4610$ & He\,I   & $2$  & $-76\pm140$ \\
             &             & Si\,III & $3$  & $-103\pm180$ \\
             &             & Metal  & $19$ & $-86\pm80$ \\
\noalign{\smallskip}
\hline
\end{tabular}
\label{tab:mag}
\end{center}
\end{table}

\section{Is HD\,76431 a binary?}

Differential radial velocities of HD\,76431 have been measured by cross-correlating the individual spectra. We detected a shift of $10\,{\rm km\,s^{-1}}$ within one day and a maximum shift of $27\,{\rm km\,s^{-1}}$ within $56$ days indicating that HD\,76431 is in a close binary system. However, the RV measurements given by \citet{ramspeck01} and \citet{behr03} are in perfect agreement ($47\pm2\,{\rm km\,s^{-1}}$ and $46.9\pm0.7\,{\rm km\,s^{-1}}$, respectively). In order to check this we gathered unpublished RV measurements provided by Green, Barlow and Edelmann (all priv. comm.). All these measurements are consistent with the literature values. A companion, if present, may therefore reside in a rather wide and eccentric orbit.

Light curves of this star have been taken by the planetary transit surveys ASAS \citep{pojmanski97} and NSVS \citep{wozniak04}. No obvious variations indicative of a close companion are present in these data. \citet{thejll95} searched for IR-excesses in the SEDs of hot subdwarf stars due to the presence of unresolved cool companions. No excess was detected for HD\,76431. Whether this star has a companion remains unclear up to now. If so, the companion must be an object of low luminosity compared to the subdwarf, most likely a late main sequence star or a white dwarf.

\section{Conclusion}

In contrast to the detection reported by \citet{otoole05} no significant magnetic field could be found in HD\,76431. This result is consistent with the non-detection by \citep{elkin98}. Given that the measurement  obtained by \citet{otoole05} is correct, the magnetic field of this star may be variable on a timescale of years. The reason for this variation may be slow rotation of the star. The sharp metal lines indicate a projected rotational velocity $<10\,{\rm km\,s^{-1}}$. 

However, Petit et al. (these proceedings) also report the non-detection of a magnetic field in HD\,76431 based on high-resolution spectropolarimetry. Furthermore, the authors reanalysed the FORS1-data on which the results of \citet{otoole05} are based and detected no significant magnetic field either. They conclude that the detection of \citet{otoole05} was spurious and caused by instrumental effects. 
The question, whether the subdwarf has a companion, remains unclear.



\end{document}